\documentclass[aps,prd,twocolumn,amsmath,amssymb,amsfonts,nofootinbib,superscriptaddress]{revtex4-1}

\usepackage{graphicx}
\usepackage{dcolumn}
\usepackage{bm}

\begin{document}
\title{How effective is machine learning to detect long transient gravitational waves from neutron stars in a real search?}

\author{Andrew L. Miller}
\affiliation{INFN, Sezione di Roma, I-00185 Roma, Italy}
\affiliation{Universit\`a di Roma La Sapienza, I-00185 Roma, Italy}
\affiliation{University of Florida, Gainesville, Florida 32611, USA}

\author{Pia Astone}
\affiliation{INFN, Sezione di Roma, I-00185 Roma, Italy}
\author{Sabrina D'Antonio}
\affiliation{INFN, Sezione di Roma Tor Vergata, I-00133 Roma, Italy}
\author{Sergio Frasca}
\author{Giuseppe Intini}
\affiliation{INFN, Sezione di Roma, I-00185 Roma, Italy}
\affiliation{Universit\`a di Roma La Sapienza, I-00185 Roma, Italy}
\author{Iuri La Rosa}
\affiliation{Universit\`a di Roma La Sapienza, I-00185 Roma, Italy}
\affiliation{Laboratoire d'Annecy-le-Vieux de Physique des Particules (LAPP),Universit\'e Savoie Mont Blanc, CNRS/IN2P3, F-74941 Annecy, France}
\author{Paola Leaci}
\affiliation{INFN, Sezione di Roma, I-00185 Roma, Italy}
\affiliation{Universit\`a di Roma La Sapienza, I-00185 Roma, Italy}
\author{Simone Mastrogiovanni}
\affiliation{PC, AstroParticule et Cosmologie, Universit\'e Paris Diderot, CNRS/IN2P3, CEA/Irfu, Observatoire de Paris,
Sorbonne Paris Cit\'e, F-75205 Paris Cedex 13, France}
\author{Federico Muciaccia}
\affiliation{Universit\`a di Roma La Sapienza, I-00185 Roma, Italy}
\author{Andonis Mitidis}
\affiliation{Capstone IT, Omaha, Nebraska 68137, USA}
\author{Cristiano Palomba}
\affiliation{INFN, Sezione di Roma, I-00185 Roma, Italy}
\author{Ornella J. Piccinni}
\affiliation{INFN, Sezione di Roma, I-00185 Roma, Italy}
\affiliation{Universit\`a di Roma La Sapienza, I-00185 Roma, Italy}
\author{Akshat Singhal}
\affiliation{INFN, Sezione di Roma, I-00185 Roma, Italy}
\author{Bernard F. Whiting}
\affiliation{University of Florida, Gainesville, Florida 32611, USA}
\author{Luca Rei}
\affiliation{INFN, Sezione di Genova, I-16146, Italy}


\date{\today}

\begin{abstract}
We present a comprehensive study of the effectiveness of Convolution Neural Networks (CNNs) to detect long duration transient gravitational-wave signals lasting $O(hours-days)$ from isolated neutron stars. We determine that CNNs are robust towards signal morphologies that differ from the training set, and they do not require many training injections/data to guarantee good detection efficiency and low false alarm probability. In fact, we only need to train one CNN on signal/noise maps in a single 150 Hz band; afterwards, the CNN can distinguish signals/noise well in any band, though with different efficiencies and false alarm probabilities due to the non-stationary noise in LIGO/Virgo. We demonstrate that we can control the false alarm probability for the CNNs by selecting the optimal threshold on the outputs of the CNN, which appears to be frequency dependent. Finally we compare the detection efficiencies of the networks to a well-established algorithm, the Generalized FrequencyHough (GFH), which maps curves in the time/frequency plane to lines in a plane that relates to the initial frequency/spindown of the source. The networks have similar sensitivities to the GFH but are orders of magnitude faster to run and can detect signals to which the GFH is blind. Using the results of our analysis, we propose strategies to apply CNNs to a real search using LIGO/Virgo data to overcome the obstacles that we would encounter, such as a finite amount of training data. We then use our networks and strategies to run a real search for a remnant of GW170817, making this the first time ever that a machine learning method has been applied to search for a gravitational wave signal from an isolated neutron star.
\end{abstract}

\pacs{Valid PACS appear here}
\maketitle


\section{\label{intro}Introduction}


The multi-messenger era of gravitational-wave physics has begun. In the past three years LIGO/Virgo \cite{aligo,avirgo} has detected many binary black-hole (BBH) mergers and one binary neutron-star (BNS) merger \cite{abbott2016observation, gw170817FIRST,abbott2018gwtc}. However, BBHs and BNSs are not the only sources of gravitational waves: we expect that rapidly spinning neutron stars with a deformation induced in their structure, in which an oscillation mode is excited, could also emit gravitational waves \cite{Jaranowski:1998qm,bauswein2012measuring,Owen:1998xg}. 

In the LIGO/Virgo collaborations, many methods have been developed to search for transient signals that last $O(s)$ and for continuous waves (CWs), quasi-infinite signals from spinning neutron stars \cite{Abbott:2017mnu,pisarski2019all,riles2017recent}. For intermediate duration signals lasting $O(hours-days)$, a cross-correlation-based method was developed some years ago \cite{Thrane:2010ri,thrane2015detecting}. But now, two categories of these types of signals exist, requiring two different analysis procedures: (1) CW-like signals with still slowly varying frequencies \cite{prix2011search,keitel2019first} and (2) signals with large frequency variations. Only recently has the scientific community taken an interest in running a search for signals  in the second category \cite{longpmr}, which required the development of new methods, both modelled \cite{PhysRevD.98.102004,oliver2019adaptive} and unmodelled \cite{Sun:2018hmm}. We expect that recently born neutron stars (i.e. after a BNS merger or supernova) will emit strong gravitational radiation on this timescale \cite{Baiotti:2016qnr}. 

The rotational evolution of a spinning neutron star is best characterized by a measurable quantity called the braking index \cite{hamil2015braking}:

\begin{equation}
    n=\frac{f|\ddot{f}|}{\dot{f}^2}
    \label{nnnn}
\end{equation}
where $f$ is the rotation frequency of the neutron star, $\dot{f}$ is the spindown, and $\ddot{f}$ is the rate of change of the spindown.

Most pulsars' measured braking indices have values within the range of $n=[1.4-3]$, indicative of electromagnetic-dominated radiation \cite{archibald2016high,clark2016braking,lasky2017braking}. However, these neutron stars are significantly older than those for which we would like to search. For a young neutron star, it is possible that the emission mechanism is dominated by gravitational waves due to a large quadrupolar deformation ($n=5$) or due to small velocity/density perturbations on the neutron star (r-modes, $n=7$) \cite{Shapiro1983}.

As the LIGO/Virgo detectors become more sensitive in the coming years, we should be able to not only detect more binary neutron star mergers, but also CWs from neutron stars. The all-sky and directed searches for CW signals are computationally expensive, since one must search the entire data set in every position in the sky. Even for long duration transient signals, signals that last for $O(hours-days)$, modelled searches such as those based on the Generalized FrequencyHough (GFH) and the Adaptive Transient Hough consume weeks of computing power, especially at intermediate spindowns \cite{PhysRevD.98.102004,oliver2019adaptive}, and require that the signal follow approximately a power law. 

We therefore explore machine learning as a way to cut computation costs without losing sensitivity. The key idea of machine learning is that one gives a ``black box'' many examples of something that one wants to be characterized (digits, cats/dogs, gravitational-wave signals, etc.), and the algorithm learns how to distinguish between the different possibilities.

Neural networks have been shown to achieve extremely good sensitivities in both white and real noise, comparable with matched filtering and some unmodeled methods \cite{george2018deep,Ggeorge2018deep,gabbard2018matching,astone2018new,dreissigacker2019deep,gebhard2019convolutional}.  However, each study addresses quantitatively some but not all of the following points: (1) how to train, (2) how much to train, (3) robustness towards signals on which the networks were not trained, (4) differences in architectures' effects on detection efficiency/false alarm probability, and (5) applying the networks to a real search. Moreover, these algorithms have only focused on binary black hole signals or supernovae, and have not yet been expanded to search for long duration transient signals from isolated neutron stars.
We address these points for CNNs, developed in \cite{astone2018new}. We characterize the performance of the networks under a variety of conditions, and propose a possible design for a search using machine learning to provide triggers that would be followed up by another method capable of estimating parameters. We also compare the CNNs to an established method to search for long duration transient signals, the GFH. Finally we use the neural networks in a real search for a remnant of GW170817 and compare our results with the previous search done \cite{longpmr}.

In section \ref{sigmod}, we describe the type of signals we expect from young, isolated neutron stars, which comprise the time/frequency maps on which we train and test the CNNs. We discuss the types of networks we use and what kind of maps we give them to train/test in section \ref{methods}. In section \ref{char}, we present the major results: a comprehensive characterization of the robustness, efficiency and false alarm probability of the CNNs across many training/testing sets, including a comparison to the GFH. We then outline a possible search scheme in section \ref{search}, perform this search for a remnant of GW170817 in \ref{proof_search} and finally draw some conclusions in section \ref{concl}.

\section{\label{sigmod}Signal model}

We derive from \eqref{nnnn} the radiation emitted by an isolated neutron star will follow a power law of the form \cite{hamil2015braking}:

\begin{equation}
    \dot{f}=-k f^n,
    \label{diff_powlaws}
\end{equation}
where $k$ is a proportionality constant that relates to physical quantities depending on $n$, i.e. the dipole magnetic field for $n=3$, the ellipticity for $n=5$, and the saturation amplitude for $n=7$. As the neutron star emits energy by any of these mechanisms, it spins downs.

Integrating Eq. \eqref{diff_powlaws}, we find how the rotational frequency changes with time:

\begin{equation}
f(t)=\frac{f_0}{\left(1+k (n-1)f_0^{n-1}(t-t_0)\right)^{\frac{1}{n-1}}}
\label{powlaws}
\end{equation}
where $f_0$ is the rotational frequency at time $t_0$, and $t_0$ is also the time when our analysis begins.

For r-modes, we use the following relation \cite{alford2014gravitational} to calculate  the gravitational wave signal amplitude $h(t)$:

\begin{equation}
    h(t)=\sqrt{\frac{2^{15} \pi^7}{5}}\frac{G J M R^3}{c^5}\frac{\alpha}{d} f(t)^3,
    \label{rmodeh(t)}
\end{equation}
where $\alpha$ is the saturation amplitude, the amplitude at which non-linear effects stop the growth of the r-mode, $J=1.635 \times 10^{−2}$ is a dimensionless angular momentum for polytropic models, $c$ is the speed of light, $d$ is the luminosity distance to the source, and $M$ and $R$ are the mass and radius of the neutron star.

For the other braking index case tested in this paper ($n=5$), we model $h(t)\propto f^2$, given by \cite{sarin2018x}:

\begin{equation}
    h(t)=\frac{4\pi^2 G I_{zz}}{c^4}\frac{\epsilon}{d}f_0^2\left(1+\frac{t}{\tau}\right)^{\frac{2}{1-n}}
\end{equation}
where $I_{zz}$ is the z-component of the moment of inertia of the neutron star, $G$ is Newton's gravitational constant, $\epsilon$ is the ellipticity of the neutron star, and $\tau$ is the spindown time-scale, defined by:

\begin{equation}
    \tau=\frac{(f_0\pi)^{1-n}}{-k(1-n)}
\end{equation}

\section{\label{methods}Methods}



\begin{figure*}[ht!]
\centering
\begin{minipage}[b]{.4\textwidth}
\includegraphics[width=80mm]{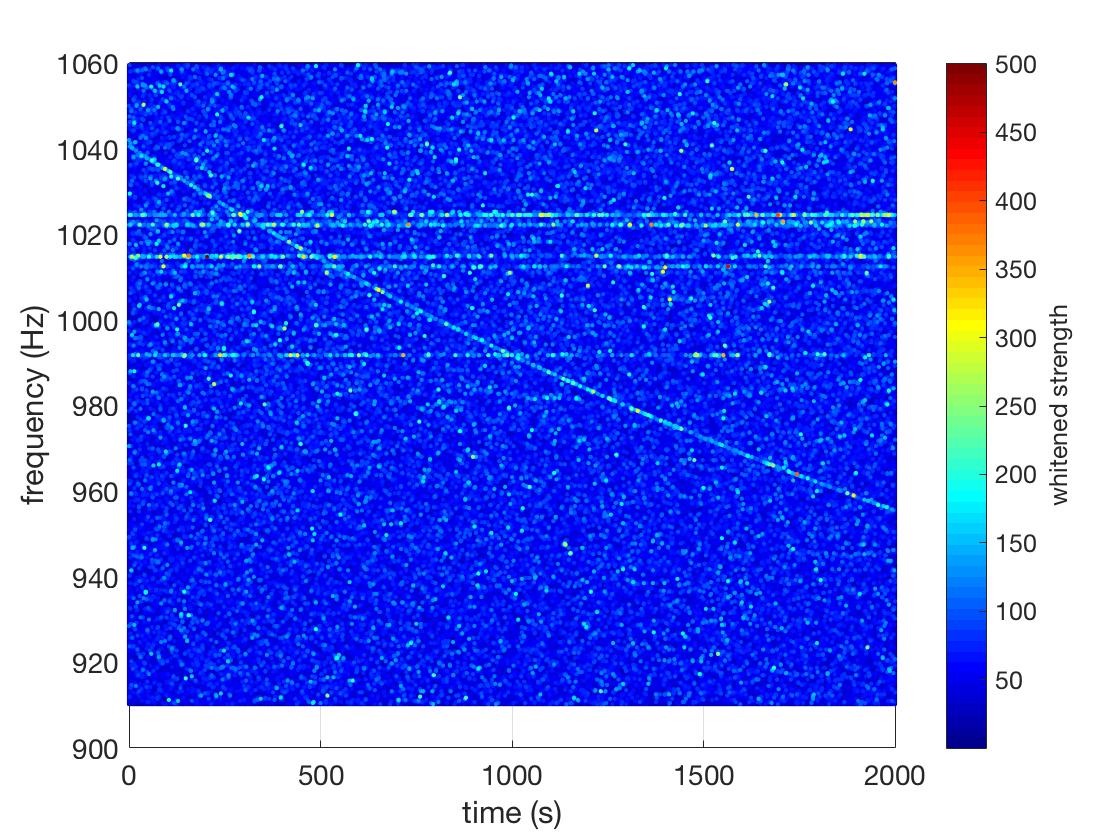}
\end{minipage}\qquad
\begin{minipage}[b]{.4\textwidth}
\includegraphics[width=80mm]{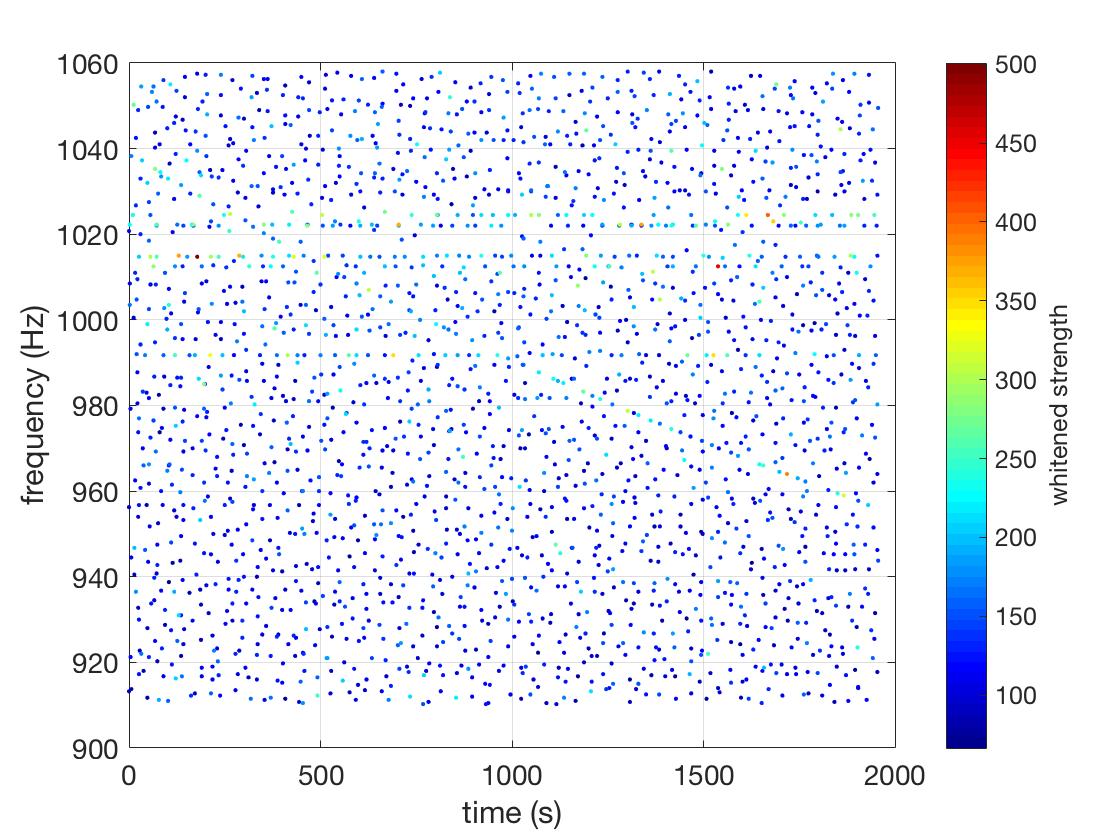}
\end{minipage}
\caption{Left: time/frequency map showing injected signal with $f_0=1040$ Hz, $h_0=7.17\times 10^{-23}$. Right: reduced resolution time/frequency map by a factor of 16 on each axis. This reduced map is the input to the machine learning algorithms. When we reduce the resolution of the image, we pick the pixel in each 16x16 block that has the highest value, but if this value is less than a threshold (2.5, essentially not higher than 2.5 standard deviations from the mean), we zero this pixel. }
\label{resredimage}
\end{figure*}

\subsection{Convolution Neural Networks}

A CNN is a biologically-inspired algorithm that uses convolutions to learn which pixels in an image are most important in distinguishing two or more classes, where in our case an ``image'' is a time/frequency map and the ``classes'' are ``map contains signal'' or ``map contains only noise''. The CNN convolves a map with a weight matrix called a ``kernel''. The kernel moves over subsets of the image and returns a measure of overlap between itself and the image, before this output is fed into an activation function. Convolutions are the mechanism by which the image is passed through each layer of the network, and allow deeper non-linear combinations of the information present in the map than Artificial Neural Networks (ANNs), where a simple matrix dot product is used to push the images through each layer \cite{gabbard2018matching,Ggeorge2018deep,george2018deep,mytidis2015sensitivity}. 

We use an implementation of CNNs shown to work for both supernovae \cite{astone2018new} and CW signals \cite{muciaccia2017using}. The key aspects of this CNN are 5 convolution blocks, described in detail below, each containing zeropadding, convolutions, ReLU and maxpooling steps, and 2 fully-connected and dropout layers followed by a softmax layer at the end.

  

The zeropadding layer ensures that the size of the image stays constant throughout the convolution block by surrounding the images with zeros. This is necessary because the maximum pooling layer takes the maximum value of every non-overlapping 2x2 block of the image, effectively cutting the size of the image in half each time.

The activation function to pass between hidden layers within the convolution block is a Rectified Linear Unit (ReLU), which simply takes either 0 or the output of the convolution $x$:

\begin{equation}
    f(x)=max(0,x)
    \label{relu}
\end{equation}
where $f$ is either $0$ or $x$.
In the fully connected layers, each output of the previous layer is used as in input to each of the 16 nodes in this layer. These layers are essentially what each hidden layer in an ANN is. The last fully-connected layer in our CNN outputs a vector whose size is just the number of classes in which we are trying to classify: 2 in this paper (signals or noise). The values correspond to the CNNs' confidence in classification of the image. 

The dropout layer is used to prevent overfitting of the network. It essentially zeroes the activation of randomly chosen hidden nodes of the model at each iteration of the learning phase, akin to pruning a decision tree.

To convert the ``confidence'' of the CNN into a probability distribution $P_i$, the final layer is a softmax activation function:

\begin{equation}
    P_i=\frac{e^{x_i}}{\sum_{n=1}^{N}e^{x_n}},
    \label{softmax}
\end{equation}
where $N$ is the number of classes (signal or noise) and $x_i$ are the outputs of the fully-connected layer (essentially the result of convolutions of the input image with the kernel).








\subsection{Construction of reduced time/frequency maps}

The behavior on which we train the networks is given in Eq. \ref{powlaws}, so we first construct a time/frequency map by fast Fourier transforming each chunk of data (interlaced by half and Tukey-windowed). Then, we estimate the noise power spectral density by using an auto-regressive spectrum to whiten the data \cite{sfdb_paper}. Afterwards we plot the result in the time/frequency plane, as shown in the left panel of Fig. \ref{resredimage}. At this level, we must clean the maps, since the CNNs are very sensitive to noise lines. Therefore we zero one or two bins that surround known noise lines \cite{Covas:2018oik}, and veto any persistent lines in the time/frequency map. This very powerful persistency veto is performed by histogramming the peakmap and projecting it onto the frequency axis. We then compute the median and standard deviation, and if there is a significant increase (more than 1 standard deviation from the median) in the number of peaks in a given frequency bin, we zero all values. This helps to decrease the false alarm probability substantially, while leaving the false dismissal probability unchanged.

Ideally, the more information that we can give to the networks, the better they can learn to identify and distinguish signals from noise. However, there are too many pixels on which to attempt to train the CNNs, so we reduce the resolution of each cleaned time/frequency map in an analogous way as done in \cite{mytidis2015sensitivity}. We split the time/frequency map into different square blocks and take only the maximum value that appears in each 16 x 16 block, if the value is above a threshold of 2.5; otherwise, we set this value to 0. The size of this box was chosen based on an optimal resolution reduction factor determined in \cite{mytidis2015sensitivity}. The value 2.5 is equivalent to selecting peaks in the spectrum that are at least 2.5 standard deviations away from the mean level of noise in the map. Selecting a local max and applying a threshold gives us the reduced time/frequency map, constructed in a similar way as  the peakmap, a collection of time/frequency peaks that are local maxes above a threshold of 2.5, in \cite{Astone:2014esa}. But our method of resolution reduction ensures that our maps have the same size, which is necessary for the machine learning to work well. In Fig. \ref{resredimage} we show the before and after resolution-reduction time/frequency maps. From here on, the term ``time/frequency maps'' will refer to the ``reduced time/frequency maps''.

\section{\label{char}Results: Machine learning characterization}

For all of these results, we have trained and tested on power-law signals in the time/frequency maps (equation \eqref{powlaws}) lasting for $2000$ s with $n=5,7$, $f_0=[250,1800]$ Hz, $\dot{f_0}=[1/164,1/16]$ Hz/s, and $h_0=[10^{-23.5},10^{-21.8}]$. All results in this section have been obtained in real O2 Livingston data from science mode periods between 4 January 2017 and 30 March 2017 \cite{goetzz}, analyzed previously in \cite{abbott2019all}, using 64000 signal maps and the same number of noise maps generated from Short Fast Fourier Databases \cite{T1900140,sfdb_paper}, though there are only 2000 unique noise maps. Any compact binary system within this time (e.g. GW170104) would have its power spread across many frequency bins, since the spinup of the signal is much greater than the resolution of a frequency bin.

\subsection{Training efficiency}
\label{efftrainsec}

\begin{figure}[ht!]
    \centering
    \includegraphics[width=\columnwidth]{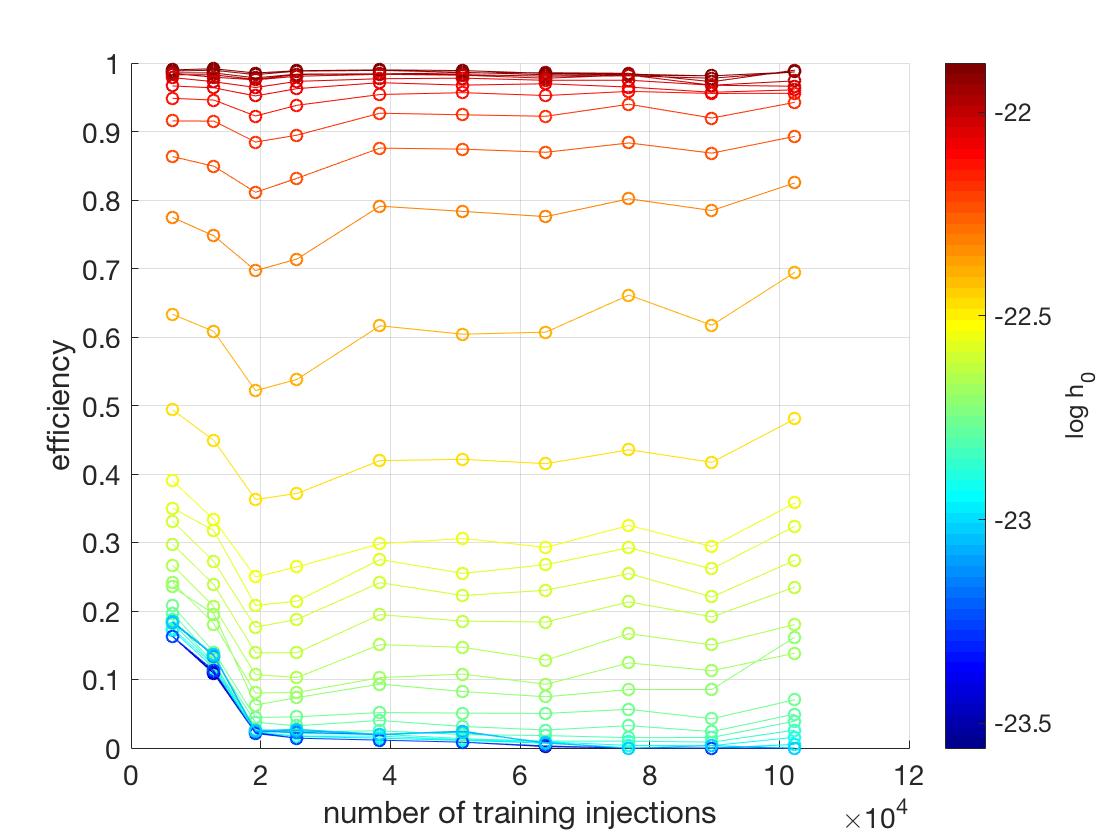}
     \caption{A CNN was trained on a variety of amplitudes in real noise with a different number of signal injections. It appears that the detection efficiency does not change much as the number of training injections is increased. Therefore, it is not necessary to do many injections to achieve high detection efficiencies. For this plot, $p_{thr}= 0.9$. Each of 32 amplitudes is represented by a line in the above figure.}
    \label{train_eff_ninjs}
\end{figure}

To run a real search, we need to train the CNNs to recognize both signals and noise. However, when we train on noise maps, we have instructed the networks to see these time/frequency maps as not having signals. We would therefore have to throw out this data, so we wish to minimize how much data we need to use for training. Thus we quantify at what point additional training of CNNs no longer adds appreciably to the detection efficiency. Fig. \ref{train_eff_ninjs} shows a CNN trained with different numbers of injections and noise maps on real data. We apply a threshold on the probability $p_{thr}$ outputted by Eq. \eqref{softmax}: $p_{thr}=0.9$, chosen based on the false alarm probability analysis done in section \ref{ddddd} . Note that $p_{thr}=0.9$ does not imply a detection efficiency of 90\%; rather, $p_{thr}$ is our detection statistic. Even for training on $\sim 20000$ injections, we see that the detection efficiencies are comparable to training on $\sim100000$ injections. This appears to be independent of signal strength. However, when training on fewer injections, we are in fact training on fewer noise maps, which means the false alarm probability is higher. These results depend on which $p_{thr}$ we use, but they offer a guideline for training a single network on a variety of amplitudes.

\subsection{Robustness towards signals not trained on}

\begin{figure*}[ht!]
\centering
\begin{minipage}[b]{.4\textwidth}
\includegraphics[width=80mm]{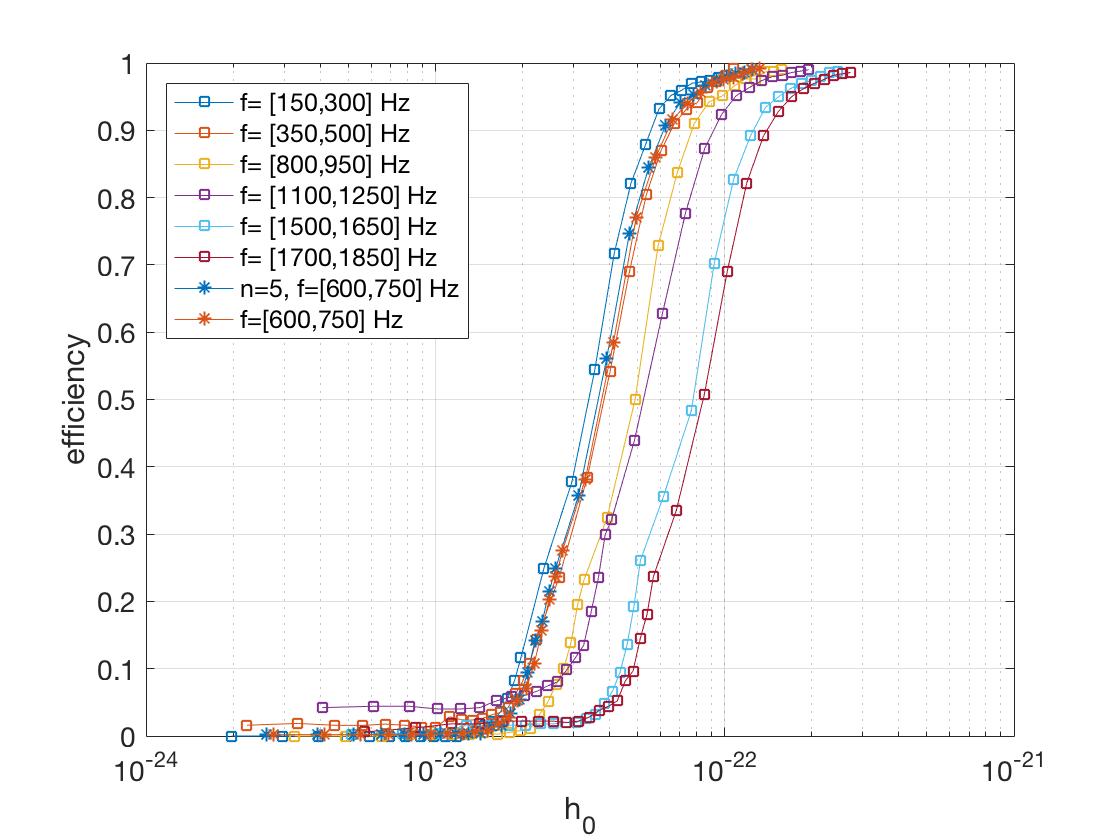}
\end{minipage}\qquad
\begin{minipage}[b]{.4\textwidth}
  \includegraphics[width=80mm]{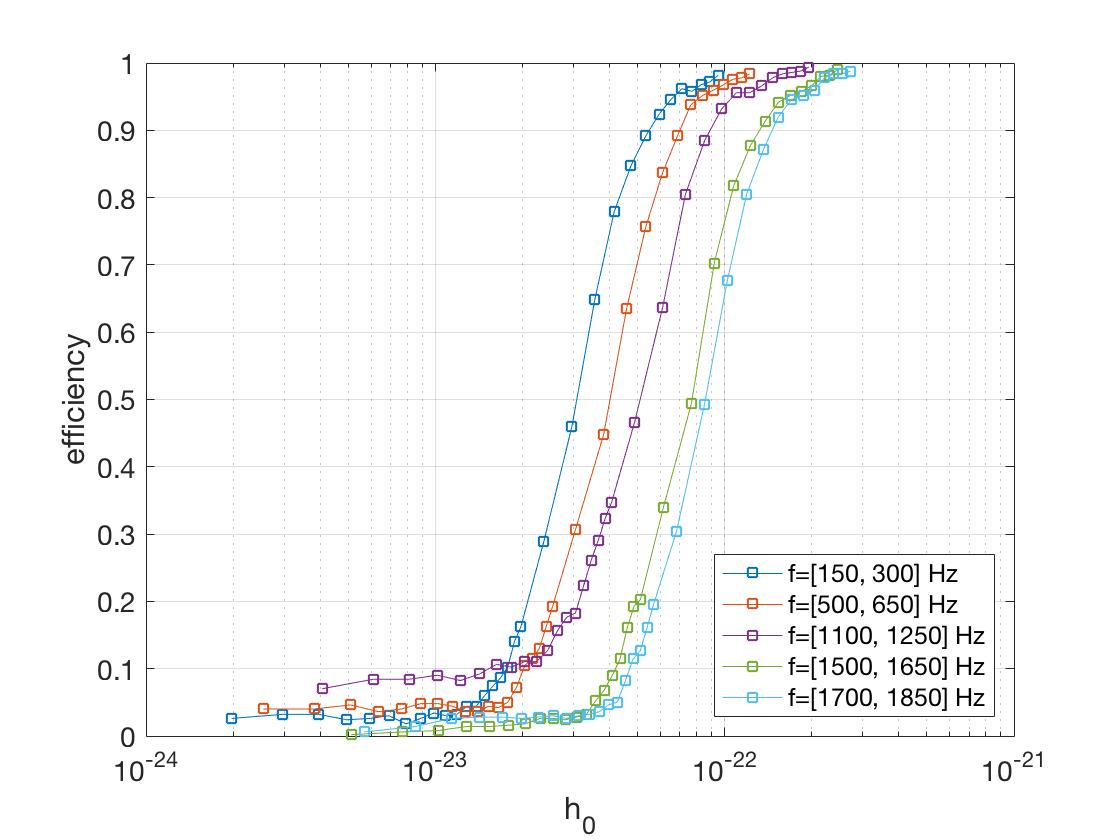}
\end{minipage}
\caption{A CNN was trained on injections with $n=7$ and tested on 16000 (500 injections/amplitude) with fixed braking indices (left) and varying braking indices (right) in different frequency bands. The random variation in braking index is between $\delta n/\delta t=[-10^{-4},10^{-4}] $Hz/s. This variation over the observation time is too large for the GFH to recover a signal. A threshold on probability $p_{thr}=0.9$ was used in both plots. The sensitivity curves for the different frequency bands are consistent with the detector's sensitivity curve (we are less sensitive at higher frequencies). The CNNs can obtain comparable sensitivities in both the fixed and varying braking index cases. However, even after cleaning both known and unknown noise lines, the threshold on probability must be high for some frequency bands to achieve a false alarm probability of $1\%$- see Fig. \ref{far_thresh}.}
\label{var_n_sens}
\end{figure*}

Machine learning methods have become useful because they can be trained to identify certain patterns, and can also generalize to find similar ones, but we must quantify if and how the detection efficiencies will change. The signals we have described thus far have had constant braking indices, which means that the GFH was a strong method to search for them. However, if the braking index varies too quickly in time, the GFH cannot recover the signal, no matter how strong the signal is, because the signal deviates too much from the power law model in Eq. \eqref{powlaws}. Fig. \ref{var_n_sens} shows two sets of sensitivity curves for a CNN trained on signals with a braking index of $n=7$. In the left panel, the curves are the result of testing on injections with fixed braking indices ($n=5$ or $n=7$). The CNNs performed worse at higher frequencies, which is consistent with the sensitivity curve of the LIGO detector. In the right panel, the CNNs were tested on injections with varying braking indices, with variation $\delta n/\delta t$ between $[-10^{-4},10^{-4}]$ /s. In total, this variation corresponds to a change $\Delta n=T_{obs}\frac{\delta n}{\delta t}=0.1$: the GFH cannot detect signals with such large $\Delta n$. Comparing both plots in Fig. \ref{var_n_sens}, we see that despite being trained only $n=7$ in the $[600,750]$ Hz band, the CNNs can detect signals with or without large $\delta n/\delta t$ and have comparable efficiencies across all frequency bands.

\subsection{Impact of $p_{thr}$ on false alarm probability}\label{ddddd}

\begin{figure}[ht!]
     \centering
     \includegraphics[width=\columnwidth]{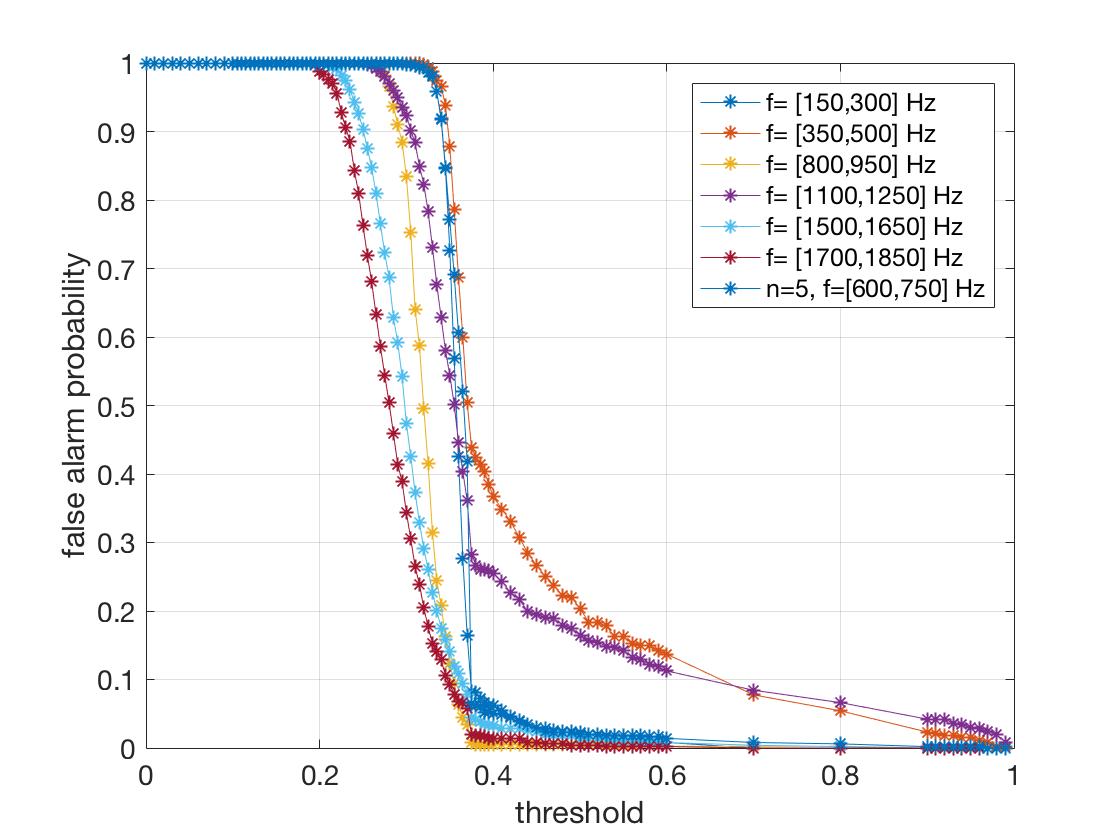}
      \caption{We plot the false alarm probability as a function of the threshold on the output of a CNN (Eq. \eqref{softmax}) for 500 noise maps per frequency band, which correspond to $\sim11.5$ days of new data. The CNN was trained on 2000 noise maps (23 days of data). Different frequency bands, which have different levels of noise, require different $p_{thr}$ to achieve the same false alarm probabilities. }
\label{far_thresh} 
\end{figure}

The output of the CNN can be interpreted as a probability that a time/frequency map contains a signal or not. If we wish to decrease the false alarm probability, we can use a threshold that is higher than $p_{thr}=0.5$. To determine an optimal threshold, we test our network on 500 noise maps generated at random times. As shown in Fig. \ref{far_thresh}, to guarantee a false alarm probability of $1\%$, we should have $p_{thr}=[0.7,0.95]$ depending on the frequency band. 
We also varied the number of noise maps used for training, and found that $\sim$ 23 days of data (2000 noise maps) was enough to obtain a false alarm probability of $\sim 1\%$ at $p_{thr}=0.9$.


\subsection{Error analysis}

\begin{figure*}[ht!]
\centering
\begin{minipage}[b]{.4\textwidth}
\includegraphics[width=80mm]{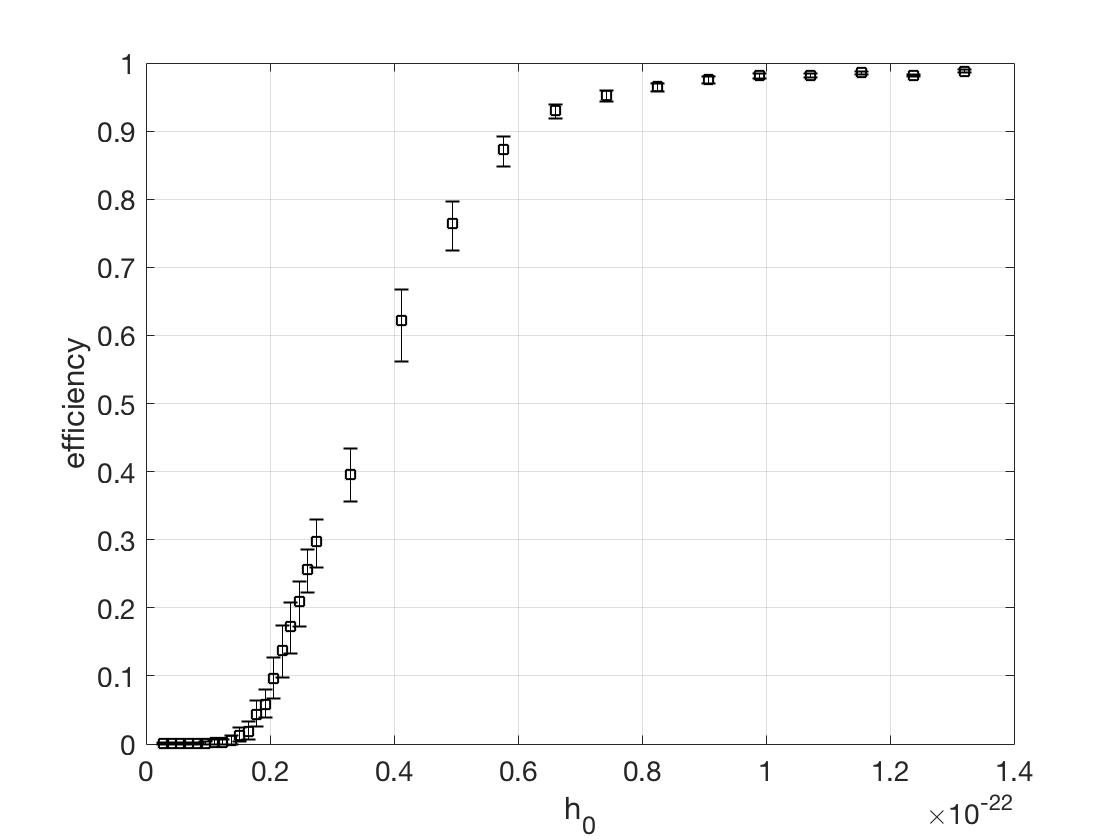}
\end{minipage}\qquad
\begin{minipage}[b]{.4\textwidth}
  \includegraphics[width=80mm]{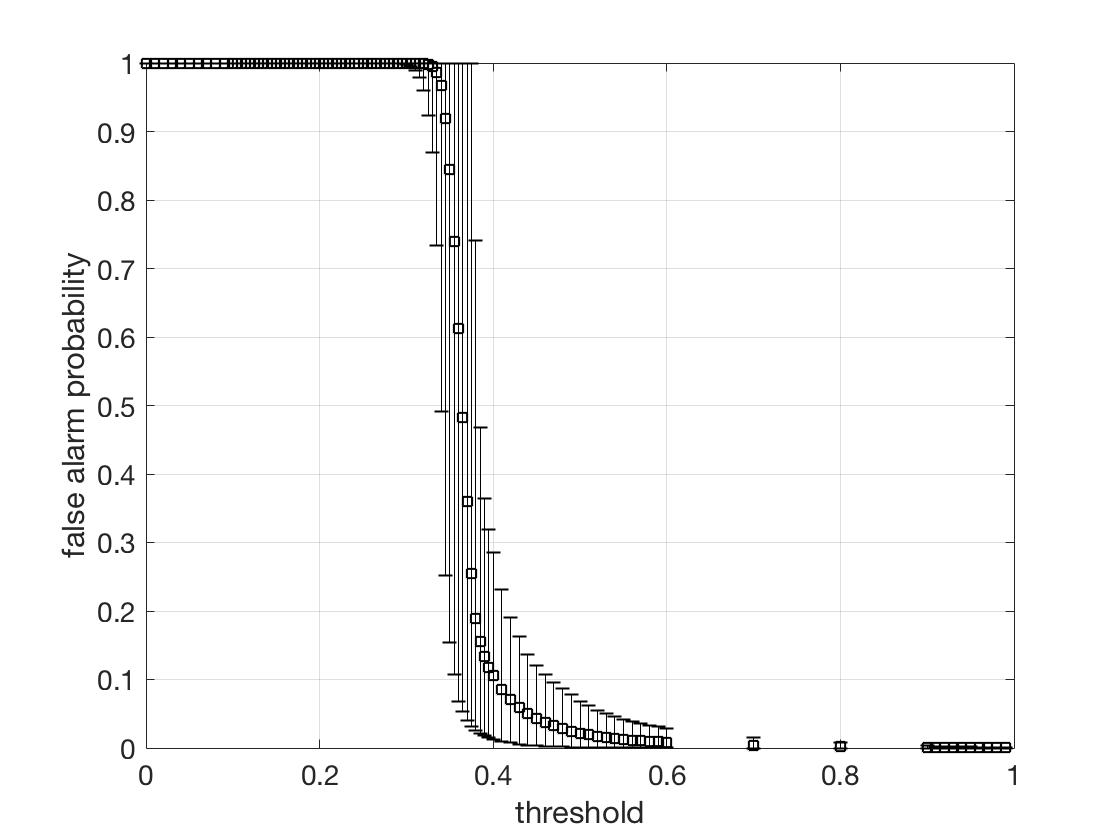}
\end{minipage}
\caption{1255 CNNs were trained on 38400 injections with $n=7$ and 1200 noise maps and tested on 44800 (1400 injections/amplitude) and 1400 noise maps. The error bars represent the 90\% confidence level. Left: efficiency is plotted as a function of amplitude. Intermediate strength signals seem to have the greatest spread in error. Right: false alarm probabilty is plotted as a function of threshold. The errors are largest around 0.4, and show that higher thresholds are necessary to truly guarantee low ($\sim 1\%)$ false alarm probabilities. }
\label{error}
\end{figure*}

The CNNs operate as a black box: we control the input, the structure of the network and the output. Because of this, there is a good degree of randomness in the possible outputs of the network, meaning that each measurement we make has an associated error. This randomness is caused by the dropout layer, which throws away half of the nodes in the network, and the fact that the network sees randomly shuffled data in different batches in each training session. We therefore train a network 1255 times with the same input data in order to quantify the range of measurements of both detection efficiency and false alarm probability. We discovered that the errors do not obey a binomial distribution for $p_{thr}=0.9$. For the [600, 750] Hz frequency band, we train on 38400 injections and 1200 noise maps. The sensitivity and false alarm probability curves are shown in figure \ref{error}.

\subsection{Comparison to Generalized FrequencyHough}

\begin{figure}[ht!]
     \centering
     \includegraphics[width=\columnwidth]{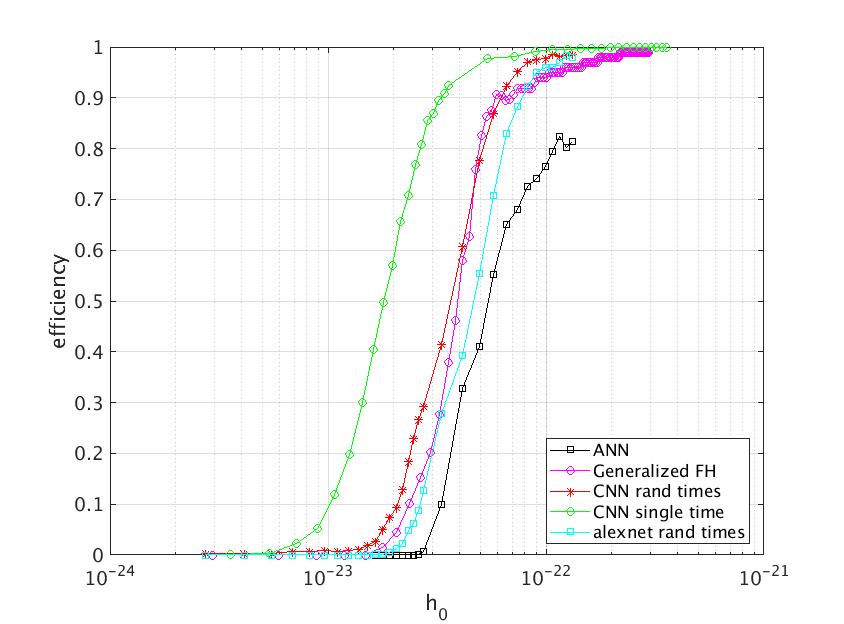}
      \caption{We show sensitivity curves for an ANN, the GFH, a CNN and Alexnet CNN trained on $2000$ injections/amplitude and tested on $1000$ injections/amplitude in real noise. $p_{thr}=0.9$. It appears that different networks obtain similar efficiencies to the GFH, though the false alarm probabilities vary: $\sim 1\%$ for the ANN/CNNs, and $\sim 0.01\%$ for the GFH. The green curve, corresponding to a CNN trained and tested at a fixed GPS time, represents what the efficiency would be if the noise was the same throughout the run. }
\label{white_real_hough_comp} 
\end{figure}

We have compared the detection efficiencies of our CNNs to those of the GFH \cite{PhysRevD.98.102004}, to Alexnet CNNs and to ANNs, developed in \cite{mytidis2015sensitivity}. In Fig. \ref{white_real_hough_comp}, we plot the sensitivity curves for each of these four methods, where we have used $p_{thr}=0.9$ for the ANNs and both CNNs. The ANN performs the worst, not even reaching 90\% efficiency, while both architectures of CNNs and the GFH achieve similar efficiencies. The CNNs are even slightly more efficient than the GFH at some amplitudes. However, the CNNs cannot perform parameter estimation and the false alarm probabilities are about 100 times greater for both CNNs than for the GFH, see Fig. \ref{far_thresh}. A ``detection'' for the GFH is defined as when the parameters $x_0=1/f_0^{n-1}$ and $k$ of the loudest candidate in the map are within a distance of 3 bins of the injected signal. For the CNNs, a detection occurs when the returned probability is greater than  $p_{thr}=0.9$. 

We also show in Fig. \ref{white_real_hough_comp} the ideal efficiency of a CNN (green curve). This is the best-case scenario because we train and test the CNN at a fixed GPS time, meaning that the CNN learns perfectly the noise structure and hence the false alarm probability is 0. The green curve represents essentially an upper limit on the CNN's sensitivity. In contrast, the black curve corresponds to the realistic case, where the CNN is affected by the non-stationary noise of the LIGO/Virgo detectors. This result shows that we must be aware of the changing noise in the detector when determining how efficient a CNN will be in detecting power-law signals. 


\section{\label{search} Search design}

We plan to use the CNNs in a real search in order not only to speed up the analysis, but also to probe a portion of the parameter space to which the GFH is not sensitive. It seems that the CNNs are of comparable efficiency to the GFH, so they can be used to quickly analyze time/frequency maps to determine if a signal is present or not. CNNs can analyze a time/frequency map in a microsecond, while the FH and GFH could take $\sim 1$ minute \cite{Astone:2014esa,PhysRevD.98.102004}. If the map contains a signal, we can then apply the GFH to estimate the parameters of the signal. Moreover, the CNNs are robust towards varying braking indices and different frequency bands with a limited amount of training data. This means that the CNNs can be trained on a small frequency band/time range and then tested across many different frequencies/times. 

Once a CNN is trained on $\sim 20000$ injections (at different amplitudes) across 150 Hz and 2000 s maps, for about 22 days of data for a false alarm probability of $1\%$ (which is about 2000 noise maps), we can apply the following procedure for each detector:

\begin{enumerate}
    \item Load short fast Fourier transform Databases \cite{sfdb_paper}.
    \item Choose appropriate FFT length depending on maximum $\dot{f_0}$, create complete time/frequency map and peakmap \cite{Astone:2014esa}.
    \item Clean known noise lines and persistent lines.
   \item Reduce resolution by a factor of 16 on each axis.
    \item Run CNNs on each reduced time/frequency map.
    \item Repeat for each detector's data. Do coincidences between time/frequency maps that produce triggers.
    \item If a trigger remains after the coincidence step, run GFH on the peakmap and obtain parameters of the signal.
    \item Perform coincidences between candidates returned by the GFH, follow-up, etc. using methodology described in \cite{PhysRevD.98.102004}.
    \item Repeat for each 150 Hz/2000 s map.
\end{enumerate}

We must emphasize that if a coincidence occurs between time/frequency maps at the level of the CNNs, the GFH can estimate the parameters of a signal only if it follows a power-law behavior. Otherwise, we can only constrain that the variation in braking index is no more than a certain $\delta n$.

As stated in section \ref{efftrainsec}, we cannot test a CNN on the same noise maps on which we train, since the CNN has learned that those maps do not contain a signal. But, if we create different CNNs that are trained and tested on different portions of the data, we can avoid this problem. The logic is the following: suppose we have data for 90 days, and that we need to train a CNN on only 2000 noise maps (23 days of data) to guarantee reasonable false alarm probability/detection efficiency comparable to the GFH. At a minimum, we can train on the first 23 days, and then run an analysis on the last 67 days. Afterwards, we could train on days 24-47, then analyze days 1-23 and 48-90. This would then require the same number of injections, just a doubling of the training time, which takes $\sim 30$ minutes. The more we repeat this ``shifting'', the more times that a different CNN will see the same data.

Since we have shown that the CNNs are robust towards signals in different frequency bands, we should not need injections in each frequency band; rather, it would be sufficient to just train on noise maps across a wide range of frequencies and injection maps in one frequency band. This greatly decreases the computational time required for such an ensemble.


\section{\label{proof_search}Search for a remnant of GW170817}

To test our procedure, we ran a search on one week of O2 Livingston and Hanford data after GW170817 to look for a remnant. We split up the search duration in 2000 s time/frequency maps, following the design outlined in the previous section.

Performing this search over one week of data spanning [100,1900] Hz took about 30 minutes. The CNNs took about 2 minutes to analyze the full data set and to generate triggers. Then, the follow-up of these 50 triggers by the GFH took about 15 minutes, when running this follow-up on 14 cores at CNAF-INFN computing center. Finally, the follow-up for the candidate remaining after the GFH took less than a minute. 

No significant candidates were found, and the upper limits are consistent with those found in \cite{longpmr}.

\subsection{Parameter space explored}

\begin{figure}[ht!]
    \centering
    \includegraphics[width=\columnwidth]{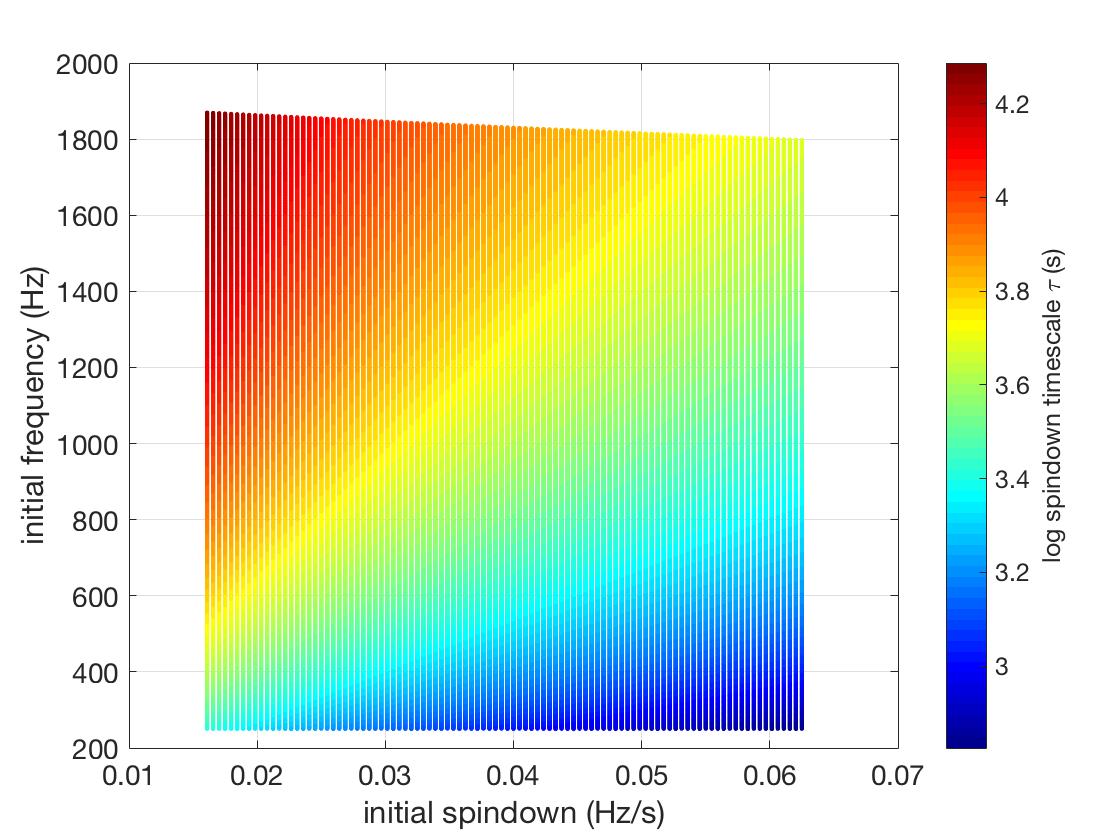}
     \caption{Parameter space explored in our search for a remnant of GW170817, for $n=7$, invoking the criterion that a signal must be fully contained within a frequency band analyzed to be detectable.}
    \label{parmspace}
\end{figure}

We model the frequency evolution of the signal as a power-law. The initial frequency and spin-down are selected so that a signal will not spin out of the analyzed frequency band during the analysis time. Fig. \ref{parmspace} plots the initial frequencies, spindowns and spindown timescales which meet this criterion for our search configuration, fixing $n=7$ (in general this plot can be created for any braking index). The spindown timescale is colored.

\subsection{Description of search results}

Coincidences done in time between the Livingston and Hanford maps analyzed by the CNNs resulted in 50 time/frequency maps having a value of $p>p_{thr}=0.9$. A grid over $n$ was constructed \cite{PhysRevD.98.102004} for each of these maps from $n=[2.5,7]$, in each 150 Hz band between $f=[100,1900]$ Hz. We then applied the GFH individually to the peakmaps generated at these times (not the maps that the network saw, but the maps constructed by the procedure in \cite{Astone:2014esa}). The GFH returned 431 coincident candidates, where a coincidence occurs if the Euclidean distance between candidate parameters $x_0/k$ in each detector is less than 3 bins. Most candidates had a very low detection statistic, called the critical ratio (CR), defined as:

\begin{equation}
    CR=\frac{y-\bar{x}}{\sigma}
\end{equation}
where $y$ is the number count returned by the GFH, $\bar{x}$ and $\sigma$ are the mean and standard deviation of the full Hough map, respectively.

We required $CR>3.5$, corresponding to a false alarm probability of $0.02\%$. After applying this threshold, 1 candidate remained (for $T_{FFT}=4$ s). We then use the full Band-Sampled-Data (BSD) \cite{piccinibsd} follow-up procedure described in \cite{PhysRevD.98.102004}. The procedure corrects the peakmaps for the time/frequency evolution of the signal, resulting (theoretically) in a monochromatic signal. $T_{FFT}$ is then increased to improve the signal-to-noise ratio and the original FrequencyHough is applied. If the critical ratio of a candidate is less now than before the follow-up, the candidate is vetoed. In each iteration, $T_{FFT}$ is doubled. The candidate remained for a few iterations until we used a $T_{FFT}=64$ s, when it was finally vetoed. We therefore conclude that there was no detectable gravitational wave emission from a remnant of GW170817.

\subsection{Upper limits}

\begin{figure}[ht!]
    \centering
    \includegraphics[width=\columnwidth]{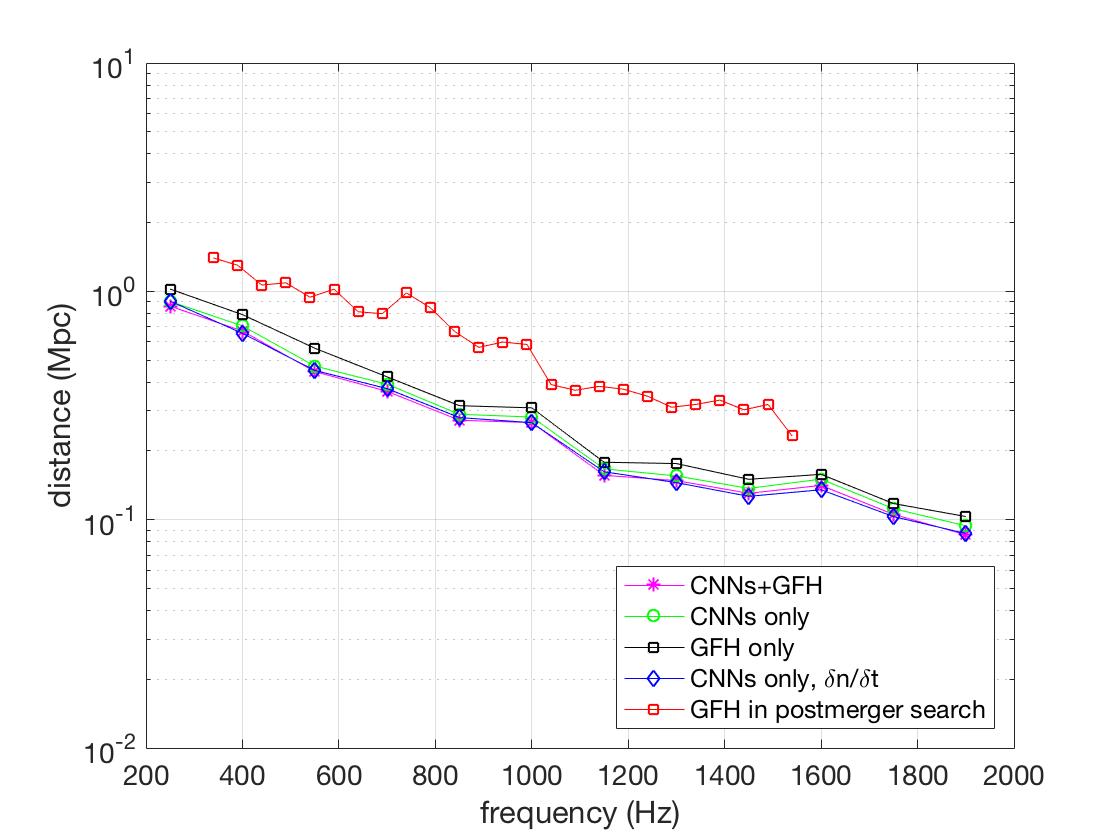}
     \caption{Upper limits on distance reach at $50\%$ confidence for the search design presented in this paper, CNNs followed up by the GFH (magenta curve) and only CNNs trying to detect time-varying braking index signals (blue). For comparison curves obtained by using just the CNNs (green), the GFH in this search (black) and the previous postmerger search (red) \cite{longpmr} are shown. The CNNs are sensitive to time-varying braking indices as well, even though these upper limits are done on fixed braking indices. Reasons why the GFH performed better in the first postmerger search include the use of a fixed $n=5$, boosting sensitivity at low frequencies, and longer duration signals at high frequencies, improving sensitivity there \cite{PhysRevD.98.102004}. Additionally the GFH in this search performs worse at higher frequency because the grid in $k$ is much finer and for signal durations of 2000 s, the signal is split among neighboring $k$ bins. Alternative ways to optimally construct the grid in $k$ are currently being investigated.}
    \label{uls}
\end{figure}

We determine the weakest signals we can see by performing 250 injections, each lasting $2000$ s, at different amplitudes with parameters uniformly distributed between $n=[2.5,7]$, $f_0=[250,1900]$ Hz, $\dot{f}=[1/64, 1/16]$ Hz/s, $\cos\iota=[-1,1]$, polarization angle $\psi=[-90,90]$ degrees, throughout the entire week after the merger. Another set of injections was also done with $\delta n/\delta t=[-10^{-4},10^{-4}]$ /s, analyzed only by the CNNs. A detection is obtained if a candidate returned by the GFH is within 3 bins of the injections and if its $CR>3.5$, the requirement used in the follow-up in the search. Fig. \ref{uls} reports upper limits on distance on a possible power-law signal coming from a remnant of GW170817 at the 50$\%$ confidence level. We have plotted five curves: one using the CNNs followed up by the GFH to estimate the parameters of the signal (magenta), one curve using just the CNNs to detect time-varying braking index signals (blue), and three other curves for comparison purposes. The red one is from the original search \cite{longpmr}, adjusted to be at the $50\%$ confidence level, and the other two curves were obtained by just using the CNNs (green) and GFH (black) to calculate the upper limits. It seems that the previous upper limits are better because the signals were simulated for 2000-16000 s consistent with the search length at $n=5$ \cite{longpmr}. This should imply an increase of at most $40\%$ in distance reach at higher frequencies, and $n=5$ corresponds to the best sensitivity by about $20\%$ relative to other braking indices at lower frequencies for the GFH, determined by using equation 35 and looking at figure 15 in \cite{PhysRevD.98.102004}. Additionally the grid in $k$ constructed during the GFH is not optimal for signals of such a short duration, especially at high frequencies, so we are currently investigating ways to improve this grid.

For each curve, we determined the distance reach by assuming a moment of inertia of $I_{zz}=100 M_{sun}^3 G^2/c^4\approx 4.34\times 10^{38}$ kg$\cdot$ m$^2$, and calculating the maximum allowed ellipticity $\epsilon$ by energy conservation, as was done in \cite{longpmr}.

\section{\label{concl}Conclusions}

We have studied how much we can trust CNNs to detect a gravitational wave signal from isolated neutron stars in real noise. For each data set that a network is given, it is necessary to study how and how much to train, the robustness of the network towards images on which it was not trained, and detection efficiency compared to a traditional algorithm. We have explored the limits of the number of training examples that could be used to train a network, and show that it is possible to achieve comparable detection efficiencies with the limited amount of LIGO/Virgo data we have and expect to obtain in the future, and at a much faster rate than traditional algorithms. We have also successfully applied machine learning to a search of real gravitational wave for a remnant of GW170817, which is the first ever search using CNNs.

In summary:

\begin{enumerate}
    \item CNNs have comparable efficiencies to the GFH but with a larger false alarm probability depending on the frequency band and how many noise maps we train on.
    \item CNNS are at least 5 orders of magnitudes faster than the GFH.
    \item Not much training data is necessary to achieve good detection efficiencies. CNNs can therefore be applied quickly to new data in a low-latency way without discarding too much data. 
    \item Training does not have to cover exhaustively the parameter space $n$=[2.5,7], $f_0$=[20, 2000] Hz, but we must have an understanding of which value to place on $p_{thr}$ since the false alarm probability may vary in each band.
    \item The CNNs are extremely sensitive to lines in the data, to the point where the CNNs will believe that every map that contains a line is a signal, implying a false alarm probability of one. It is therefore crucial to clean the time/frequency maps before giving them to the CNNs. 
    \item Complicated network architectures do not seem to be needed for this pattern recognition problem. This implies less time necessary for training and less hyperparameters to tune.
    \item The place of machine learning in long duration transient analysis is to provide triggers that can be followed up by the GFH, therefore a (not too high, around $1-3\%$) false alarm probability is not a problem.
\end{enumerate}

Future work will include training on previous observing run's data and testing on new data, in order to avoid using ensembles of CNNs. However, we expect that this will work only when the noise in each observing run does not change too much, which should be the case in future runs. We also plan to experiment with combining two or three detectors' time/frequency maps into one image, so-called RGB synthesis, that was explored for supernovae in \cite{astone2018new}. While the efficiency will probably not improve, the false alarm probability should decrease, since noise lines (which we have seen greatly increase the false alarm probability if not removed) are typically not coincident in each detector.

We are also working to estimate $n$ and $\delta n/\delta t$ using CNNs. This would allow us to further reduce the computation time necessary for the GFH, since we would not need to loop over all possible braking indices as we did in \cite{longpmr}. Additionally, if we know $\delta n/\delta t$, we can correct the time/frequency maps for this behavior before giving them to the GFH. 

\section{Acknowledgements}
The authors gratefully acknowledge the Italian Istituto Nazionale di Fisica Nucleare (INFN), and the INFN-CNAF for provision of computational resources. 
\bibliographystyle{acm}
\bibliography{biblio.bib}

\begin{thebibliography}{10}

\bibitem{aligo}
{\sc Aasi, J., Abbott, B., Abbott, R., Abbott, T., Abernathy, M., Ackley, K.,
  Adams, C., Adams, T., Addesso, P., Adhikari, R., et~al.}
\newblock Advanced ligo.
\newblock {\em Classical and quantum gravity 32}, 7 (2015), 074001.

\bibitem{abbott2018gwtc}
{\sc Abbott, B., Abbott, R., Abbott, T., Abraham, S., Acernese, F., Ackley, K.,
  Adams, C., Adhikari, R., Adya, V., Affeldt, C., et~al.}
\newblock Gwtc-1: A gravitational-wave transient catalog of compact binary
  mergers observed by ligo and virgo during the first and second observing
  runs.
\newblock {\em arXiv preprint arXiv:1811.12907\/} (2018).

\bibitem{abbott2019all}
{\sc Abbott, B., Abbott, R., Abbott, T., Abraham, S., Acernese, F., Ackley, K.,
  Adams, C., Adhikari, R., Adya, V., Affeldt, C., et~al.}
\newblock All-sky search for continuous gravitational waves from isolated
  neutron stars using advanced ligo o2 data.
\newblock {\em Physical Review D 100}, 2 (2019), 024004.

\bibitem{abbott2016observation}
{\sc Abbott, B.~P., Abbott, R., Abbott, T., Abernathy, M., Acernese, F.,
  Ackley, K., Adams, C., Adams, T., Addesso, P., Adhikari, R., et~al.}
\newblock Observation of gravitational waves from a binary black hole merger.
\newblock {\em Physical review letters 116}, 6 (2016), 061102.

\bibitem{Abbott:2017mnu}
{\sc Abbott, B.~P., et~al.}
\newblock {All-sky Search for Periodic Gravitational Waves in the O1 LIGO
  Data}.
\newblock {\em \prd 96}, 6 (2017), 062002.

\bibitem{gw170817FIRST}
{\sc Abbott, B.~P., et~al.}
\newblock Gw170817: Observation of gravitational waves from a binary neutron
  star inspiral.
\newblock {\em Phys. Rev. Lett. 119\/} (Oct 2017), 161101.

\bibitem{longpmr}
{\sc Abbott, B.~P., et~al.}
\newblock Search for gravitational waves from a long-lived remnant of the
  binary neutron star merger {GW}170817.
\newblock {\em The Astrophysical Journal 875}, 2 (apr 2019), 160.

\bibitem{avirgo}
{\sc Acernese, F., Agathos, M., Agatsuma, K., Aisa, D., Allemandou, N.,
  Allocca, A., Amarni, J., Astone, P., Balestri, G., Ballardin, G., et~al.}
\newblock Advanced virgo: a second-generation interferometric gravitational
  wave detector.
\newblock {\em Classical and Quantum Gravity 32}, 2 (2014), 024001.

\bibitem{alford2014gravitational}
{\sc Alford, M.~G., and Schwenzer, K.}
\newblock Gravitational wave emission from oscillating millisecond pulsars.
\newblock {\em Monthly Notices of the Royal Astronomical Society 446}, 4
  (2014), 3631--3641.

\bibitem{archibald2016high}
{\sc Archibald, R., Gotthelf, E., Ferdman, R., Kaspi, V., Guillot, S.,
  Harrison, F., Keane, E., Pivovaroff, M., Stern, D., Tendulkar, S., et~al.}
\newblock A high braking index for a pulsar.
\newblock {\em The Astrophysical Journal Letters 819}, 1 (2016), L16.

\bibitem{astone2018new}
{\sc Astone, P., Cerd{\'a}-Dur{\'a}n, P., Di~Palma, I., Drago, M., Muciaccia,
  F., Palomba, C., and Ricci, F.}
\newblock New method to observe gravitational waves emitted by core collapse
  supernovae.
\newblock {\em Physical Review D 98}, 12 (2018), 122002.

\bibitem{Astone:2014esa}
{\sc Astone, P., Colla, A., D'Antonio, S., Frasca, S., and Palomba, C.}
\newblock {Method for all-sky searches of continuous gravitational wave signals
  using the frequency-Hough transform}.
\newblock {\em \prd 90}, 4 (2014), 042002.

\bibitem{T1900140}
{\sc Astone, P., et~al.}
\newblock Codes to create sfdbs and o2 used time segments.
\newblock Tech. Rep. LIGO-T1900140, 2019.

\bibitem{sfdb_paper}
{\sc Astone, P., Frasca, S., and Palomba, C.}
\newblock The short fft database and the peak map for the hierarchical search
  of periodic sources.
\newblock {\em Classical and Quantum Gravity 22}, 18 (2005), S1197.

\bibitem{Baiotti:2016qnr}
{\sc Baiotti, L., and Rezzolla, L.}
\newblock {Binary neutron star mergers: a review of Einstein’s richest
  laboratory}.
\newblock {\em Rept. Prog. Phys. 80}, 9 (2017), 096901.

\bibitem{bauswein2012measuring}
{\sc Bauswein, A., and Janka, H.-T.}
\newblock Measuring neutron-star properties via gravitational waves from
  neutron-star mergers.
\newblock {\em Physical review letters 108}, 1 (2012), 011101.

\bibitem{clark2016braking}
{\sc Clark, C., Pletsch, H., Wu, J., Guillemot, L., Camilo, F., Johnson, T.,
  Kerr, M., Allen, B., Aulbert, C., Beer, C., et~al.}
\newblock The braking index of a radio-quiet gamma-ray pulsar.
\newblock {\em The Astrophysical journal letters 832}, 1 (2016), L15.

\bibitem{Covas:2018oik}
{\sc Covas, P., et~al.}
\newblock {Identification and mitigation of narrow spectral artifacts that
  degrade searches for persistent gravitational waves in the first two
  observing runs of Advanced LIGO}.
\newblock {\em \prd 97}, 8 (2018), 082002.

\bibitem{dreissigacker2019deep}
{\sc Dreissigacker, C., Sharma, R., Messenger, C., Zhao, R., and Prix, R.}
\newblock Deep-learning continuous gravitational waves.
\newblock {\em arXiv preprint arXiv:1904.13291\/} (2019).

\bibitem{pisarski2019all}
{\sc et~al., B.~A.}
\newblock All-sky search for continuous gravitational waves from isolated
  neutron stars using advanced ligo o2 data.
\newblock {\em arXiv preprint arXiv:1903.01901\/} (2019).

\bibitem{gabbard2018matching}
{\sc Gabbard, H., Williams, M., Hayes, F., and Messenger, C.}
\newblock Matching matched filtering with deep networks for gravitational-wave
  astronomy.
\newblock {\em Physical review letters 120}, 14 (2018), 141103.

\bibitem{gebhard2019convolutional}
{\sc Gebhard, T.~D., Kilbertus, N., Harry, I., and Sch{\"o}lkopf, B.}
\newblock Convolutional neural networks: a magic bullet for gravitational-wave
  detection?
\newblock {\em arXiv preprint arXiv:1904.08693\/} (2019).

\bibitem{Ggeorge2018deep}
{\sc George, D., and Huerta, E.}
\newblock Deep learning for real-time gravitational wave detection and
  parameter estimation: Results with advanced ligo data.
\newblock {\em Physics Letters B 778\/} (2018), 64--70.

\bibitem{george2018deep}
{\sc George, D., and Huerta, E.}
\newblock Deep neural networks to enable real-time multimessenger astrophysics.
\newblock {\em Physical Review D 97}, 4 (2018), 044039.

\bibitem{goetzz}
{\sc Goetz, E., and LSC/Virgo}.
\newblock Segments used for creating standard sfts in o2 data.
\newblock {\em Tech. Rep. LIGO-T1900085\/}.

\bibitem{hamil2015braking}
{\sc Hamil, O., Stone, J.~R., Urbanec, M., and Urbancova, G.}
\newblock Braking index of isolated pulsars.
\newblock {\em Physical Review D 91}, 6 (2015), 063007.

\bibitem{Jaranowski:1998qm}
{\sc Jaranowski, P., Kr{\'o}lak, A., and Schutz, B.~F.}
\newblock {Data analysis of gravitational-wave signals from spinning neutron
  stars: The signal and its detection}.
\newblock {\em \prd 58\/} (1998), 063001.

\bibitem{keitel2019first}
{\sc Keitel, D., Woan, G., Pitkin, M., Schumacher, C., Pearlstone, B., Riles,
  K., Lyne, A.~G., Palfreyman, J., Stappers, B., and Weltevrede, P.}
\newblock First search for long-duration transient gravitational waves after
  glitches in the vela and crab pulsars.
\newblock {\em arXiv preprint arXiv:1907.04717\/} (2019).

\bibitem{lasky2017braking}
{\sc Lasky, P.~D., Leris, C., Rowlinson, A., and Glampedakis, K.}
\newblock The braking index of millisecond magnetars.
\newblock {\em The Astrophysical Journal Letters 843}, 1 (2017), L1.

\bibitem{PhysRevD.98.102004}
{\sc Miller, A., Astone, P., D'Antonio, S., Frasca, S., Intini, G., La~Rosa,
  I., Leaci, P., Mastrogiovanni, S., Muciaccia, F., Palomba, C., Piccinni,
  O.~J., Singhal, A., and Whiting, B.~F.}
\newblock Method to search for long duration gravitational wave transients from
  isolated neutron stars using the generalized frequency-hough transform.
\newblock {\em Phys. Rev. D 98\/} (Nov 2018), 102004.

\bibitem{muciaccia2017using}
{\sc Muciaccia, F.}
\newblock Using deep convolutional neural networks to build a low-latency
  classifier to trigger the detection of long gravitational wave transients.

\bibitem{mytidis2015sensitivity}
{\sc Mytidis, A., Panagopoulos, A.~A., Panagopoulos, O.~P., Miller, A., and
  Whiting, B.}
\newblock Sensitivity study using machine learning algorithms on simulated
  $r$-mode gravitational wave signals from newborn neutron stars.
\newblock {\em Phys. Rev. D 99\/} (Jan 2019), 024024.

\bibitem{oliver2019adaptive}
{\sc Oliver, M., Keitel, D., and Sintes, A.~M.}
\newblock Adaptive transient hough method for long-duration gravitational wave
  transients.
\newblock {\em Physical Review D 99}, 10 (2019), 104067.

\bibitem{Owen:1998xg}
{\sc Owen, B.~J., Lindblom, L., Cutler, C., Schutz, B.~F., Vecchio, A., and
  Andersson, N.}
\newblock {Gravitational waves from hot young rapidly rotating neutron stars}.
\newblock {\em \prd 58\/} (1998), 084020.

\bibitem{piccinibsd}
{\sc Piccinni, O., Astone, P., D’Antonio, S., Frasca, S., Intini, G., Leaci,
  P., Mastrogiovanni, S., Miller, A., Palomba, C., and Singhal, A.}
\newblock A new data analysis framework for the search of continuous
  gravitational wave signals.
\newblock {\em Classical and Quantum Gravity 36}, 1 (2018), 015008.

\bibitem{prix2011search}
{\sc Prix, R., Giampanis, S., and Messenger, C.}
\newblock Search method for long-duration gravitational-wave transients from
  neutron stars.
\newblock {\em Physical Review D 84}, 2 (2011), 023007.

\bibitem{riles2017recent}
{\sc Riles, K.}
\newblock Recent searches for continuous gravitational waves.
\newblock {\em Modern Physics Letters A 32}, 39 (2017), 1730035.

\bibitem{sarin2018x}
{\sc Sarin, N., Lasky, P.~D., Sammut, L., and Ashton, G.}
\newblock X-ray guided gravitational-wave search for binary neutron star merger
  remnants.
\newblock {\em Physical Review D 98}, 4 (2018), 043011.

\bibitem{Shapiro1983}
{\sc {Shapiro}, S.~L., and {Teukolsky}, S.~A.}
\newblock {\em {Black holes, white dwarfs, and neutron stars: The physics of
  compact objects}}.
\newblock Wiley, New York, 1983.

\bibitem{Sun:2018hmm}
{\sc Sun, L., and Melatos, A.}
\newblock Application of hidden markov model tracking to the search for
  long-duration transient gravitational waves from the remnant of the binary
  neutron star merger gw170817.
\newblock {\em Physical Review D 99}, 12 (2019), 123003.

\bibitem{Thrane:2010ri}
{\sc Thrane, E., et~al.}
\newblock {Long gravitational-wave transients and associated detection
  strategies for a network of terrestrial interferometers}.
\newblock {\em \prd 83\/} (2011), 083004.

\bibitem{thrane2015detecting}
{\sc Thrane, E., Mandic, V., and Christensen, N.}
\newblock Detecting very long-lived gravitational-wave transients lasting hours
  to weeks.
\newblock {\em Physical Review D 91}, 10 (2015), 104021.

\end{thebibliography}

\end{document}